\documentclass[pdflatex,sn-mathphys-num]{sn-jnl}
\usepackage{graphicx}
\usepackage{multirow}
\usepackage{amsmath,amssymb,amsfonts}%
\usepackage{amsthm}
\usepackage[mathcal]{eucal}
\usepackage[title]{appendix}
\usepackage{xcolor}
\newcommand{\chg}[1]{{\color{blue}#1}}
\usepackage{textcomp}
\usepackage{manyfoot}
\usepackage{booktabs}
\usepackage{array}
\usepackage{algorithm}
\usepackage{algorithmicx}
\usepackage{algpseudocode}
\usepackage{listings}
\usepackage{hyperref}
\usepackage{caption}
\usepackage{float}
\usepackage{booktabs} 
\usepackage{afterpage}
\usepackage{subcaption}
\usepackage[utf8]{inputenc}
\usepackage{amsmath, amssymb}
\usepackage{geometry}
\usepackage{newtxtext,newtxmath}
\theoremstyle{thmstyleone}%

\theoremstyle{thmstyletwo}%
\theoremstyle{thmstylethree}%
\begin{document}
\title[final version]
{
Fractional Bose–Einstein Condensate Dark Matter Cores: Numerical GPP Solutions
and SPARC Rotation Curves
}
\author[1]{\fnm{Akila} \sur{Labadi}}\email{a.labadi@univ-chlef.dz}
\author*[1]{\fnm{Mohamed} \sur{Benarous}}\email{m.benarous@univ-chlef.dz}
\author[1]{\fnm{Ahmed} \sur{Hocine}}\email{ah.hocine@univ-chlef.dz}
\equalcont{These authors contributed equally to this work.}
\affil[1]{\orgdiv{Laboratory for Theoretical Physics and Material Physics}, \orgname{Faculty of Exact Sciences and Informatics, Department of Physics }, \orgaddress{\street{Hassiba Benbouali University}, \city{ Chlef}, \postcode{02000}, \country{Algeria}}}

\abstract{
We investigate a fractional Bose--Einstein condensate dark-matter scenario in which an
ultralight bosonic component forms a core embedded in a predominantly
cold-dark-matter halo. We solve the stationary Gross--Pitaevskii--Poisson system beyond
the Thomas--Fermi approximation using an iterative relaxation procedure and confront the
resulting density profiles with the inner rotation curves of 30 SPARC galaxies. For a
representative parameter scale ($m\sim10^{-26}\,\mathrm{eV}/c^2$ and $a\sim10^{-82}\,\mathrm{cm}$), we obtain equilibrium configurations, returned by the solver with negative total energy, and
good phenomenological fits to the \chg{inner} rotation curves of the selected galaxies. The resulting solutions exhibit finite
central densities and cored profiles. \chg{In the fits only the condensate is included as a dark-matter component: no CDM envelope is modelled, and the baryons enter only through the total velocity, an approximation that we show to be justified for the boson mass considered here.} We further examine the sensitivity of the fits to the
boson mass, \chg{discuss the cosmological and dynamical constraints on such light bosons,} and discuss the limitations associated with the treatment of the outer CDM
halo and the absence of a dynamical stability analysis.

}

\maketitle
\section{Introduction}\label{introduction}
Self-gravitating Bose--Einstein condensates (BECs) lie at the intersection of quantum many-body physics and astrophysics. In the mean-field limit, a condensate of a macroscopic number of bosons is described by a single order parameter that satisfies the Gross--Pitaevskii equation (GPE)~\cite{Cd13,Cd13a,Cd14}, and the gravitational field is determined self-consistently by Poisson's equation. The Gross--Pitaevskii--Poisson (GPP) system obtained is essentially a nonlinear many-body problem reduced to its mean-field description, where the existence and stability of bound states is governed by the combined effects of quantum pressure, self-interaction, and gravity. The galactic dark matter halo problem studied in this work is one concrete example of that broader class of self-gravitating Bose systems.

One of the biggest unsolved problems in cosmology and astrophysics is the nature of dark matter (DM). DM accounts for approximately 85\% of the matter content of the Universe~\cite{Cd1}, yet it has escaped direct detection, and the evidence for its existence is almost solely based on its gravitational effects on astrophysical systems~\cite{Cd2,Cd35}. One of strongest pieces of evidence for this comes from the flat rotation curves of spiral galaxies~\cite{Cd3}. The proposed DM candidates include weakly interacting massive particles (WIMPs)~\cite{Cd4}, axions~\cite{Cd5,Cd6}, and macroscopic compact halo objects (MACHOs) or primordial black holes, which are now strongly limited by observations of gravitational-lensing~\cite{Cd7}. DM models are broadly divided into cold dark matter (CDM)~\cite{Cd8}, ultra-light dark matter (ULDM)~\cite{Cd9}, warm dark matter (WDM)~\cite{Cd10}, and self-interacting dark matter (SIDM)~\cite{Cd11}.

\chg{Self-gravitating bosonic configurations were first studied by Ruffini and Bonazzola~\cite{Cd38}.} B\"ohmer and Harko suggested that the DM halos of galaxies could be described as gravitationally self-bound BEC clouds~\cite{Cd12}. This framework directly links the microscopic properties of the condensate—particle mass, scattering length, and chemical potential—to observable galactic kinematics, and it naturally yields cored density profiles, providing a possible solution to the cusp-core problem that plagues standard $\Lambda$CDM. The underlying GPE has been extensively verified since the experimental achievement of BEC in 1995~\cite{Cd15, Cd16,Cd17}, and has since been widely extended to cosmological contexts~\cite{Cd12,Cd18,Cd19}.

The original B\"ohmer--Harko treatment used the Thomas–Fermi (TF) approximation, neglecting the kinetic-energy (quantum-pressure) term. This simplification allows for analytic or semi-analytic solutions and reproduces core-like profiles, but later work has demonstrated that it can misrepresent both structural properties and, importantly, stability. Guzm\'an et al.~\cite{Cd20} showed that the GPE in the TF regime gives rise to unstable halo solutions for an ultra-light, non-self-interacting condensate; later on, Madarassy and T\'oth~\cite{Cd21} and T\'oth~\cite{Cd22} verified numerically that such instability is an artefact of the TF approximation, and not a physical property of the condensate. \chg{Quantum-pressure effects beyond the TF limit have, however, been treated before: by the Schive et al.\ soliton fit and by Gaussian ans\"atze, which have been compared with exact solutions obtained by imaginary-time propagation~\cite{Cd33,Cd31,Bar18,Liu23} and extended to rotating haloes~\cite{Nik21,Kor23}, and, including self-interaction, by numerical studies of the full GPP system~\cite{Cd37,Cd29}. What remains to be assessed is how stationary solutions of the full GPP system with self-interaction confront galactic rotation curves for very light bosons, which is the question addressed here.}

Standard BEC-DM scenarios also face tight cosmological bounds. Without self-interaction, matching astrophysical core sizes requires $m \sim 10^{-22}\,\mathrm{eV}/c^2$, which allows the condensate's de Broglie wavelength to be visible on galactic scales~\cite{Cd23}. Much lighter masses, $m \sim 10^{-26}\,\mathrm{eV}/c^2$, are difficult to reconcile with structure-formation constraints if the condensate is assumed to make up the \emph{entire} DM budget. We \chg{attempt to alleviate} this tension by considering a \emph{fractional} BEC-DM scenario, where the condensate forms only a sub-dominant component of the halo while standard CDM governs the outskirts and large-scale structure~\cite{Cd25,Cd26}. \chg{We stress that this reduces, but does not remove, the tension with cosmological and dynamical bounds, which we discuss in Section~\ref{sec:constraints}.}

In this paper we solve the complete GPP system using a self-consistent iterative relaxation scheme that retains the full kinetic-energy term, thus incorporating quantum-pressure effects exactly within the mean-field description, thereby extending the analysis of T\'oth~\cite{Cd22}.
We take ultra-light axion-like particles with $m \simeq 10^{-26}\,\mathrm{eV}/c^2$ and $a \simeq 10^{-82}\,\mathrm{cm}$~\cite{Cd20,Cd24}, and compare the equilibrium configurations that follow from them with rotation-curve data from the SPARC database~\cite{Cd27,Cd28}. 
As detailed in Section~\ref{sec:analysis}, we show that \chg{equilibrium solutions, which the solver returns with negative total energy, can reproduce the inner galactic kinematics of the fitted galaxies and have} flat, finite-density cores \chg{consistent} with the observed inner rotation curves. \chg{The fits include only the condensate as a dark-matter component (Section~\ref{sec:density}).}

In addition to this astrophysical use, the numerical scheme itself is a general-purpose tool applicable to the broader class of self-gravitating and trapped Bose systems -- boson stars, dilute condensates in compact objects, and large-$N$ trapped atomic condensates -- for which the GPP-type equations solved here are the shared mean-field description.

The paper is structured as follows. In Section~\ref{sec:system} we present the GPP model, its non-dimensional form, the boundary conditions and the iterative numerical method. In Section~\ref{sec:analysis} we present the rotation-curve fits and stability analysis for thirty SPARC galaxies. The results are discussed and compared to alternative dark matter scenarios in Section~\ref{sec:discussion}. We summarize our conclusions in Section~\ref{sec:conclusion}.

\section{Quantum Mechanical Framework and Numerical Method}
\label{sec:system}
\subsection{The Gross-Pitaevskii-Poisson system}

The GP equation\cite{Cd12} describes the evolution of the BEC's wave function using a non-relativistic approach based on Newtonian gravity:
\begin{equation}
\imath\hbar\,\frac{\partial\Psi(r,t)}{\partial\,t} =\left(-{\frac{\hbar ^{2}}{2m}}\Delta +V_{{\rm{grav}}}(r,t)
+g|\Psi|^{2}(r,t)\right)
\Psi(r,t),
\label{eq1}   
\end{equation}
where the gravitational potential satisfies Poisson's equation:
\begin{equation}
\Delta{V_{{\rm {grav}}}(r,t)}=4\pi G m^2\, |\Psi|^{2}(r,t).
\label{eq2}
\end{equation}
In Eqs.(\ref{eq1}-\ref{eq2}), $m$ denotes the mass of the particles that constitute the condensate, and $g$ represents their coupling constant, which is linked to the s-wave scattering length $a$ through the relation $g=4\pi \hbar^2 a/m$. \\
\emph{Equilibrium Configuration.} Writing the equilibrium condensate wave function as $\Psi(r,t)=e^{-i\mu t/\hbar}\psi(r)$, with $\mu$ the chemical potential, the time-independent Gross-Pitaevskii-Poisson equations are readily obtained:
\begin{equation}
\mu\psi(r) =\left[-{\frac{\hbar ^{2}}{2m}}\Delta+V_{{\rm
{grav}}}(r)+g{\psi}^{2}(r)\right]\psi(r),
\label{eq3}
\end{equation}
\begin{equation}
\Delta V_{{\rm {grav}}}(r)=4\pi G m^2 \psi^{2}(r),
\label{eq4}
\end{equation}
where for neutral particles $\psi (r)$ can be taken real without loss of generality.

\subsection{Nondimensional Formulation} 
To solve numerically these equations, we begin by defining the dimensionless quantities: $\tilde{r}=r/R_{c}$, ${\tilde{\psi}}^{2}=\left(8\pi G m^{3}R_{c}^{4}/\hbar^{2}\right)\psi^{2}$, $\tilde{V}_{\rm{grav}}=\left(2 m R_{c}^{2}/\hbar^{2}\right)V_{\rm{grav}}$, $\tilde{a}=\left(\hbar^{2}/G m^{3}R_{c}^{2}\right)a$, and $\tilde{\mu}=\left(2 m R_{c}^{2}/\hbar^{2}\right)\mu$. $R_{c}$ is the characteristic core radius. The coupled equations (\ref{eq3}-\ref{eq4}) are then \chg{written} as
\begin{equation}
\left(-\Delta_{\tilde{r}} +\tilde{V}_{{\rm{grav}}}+\tilde{a}{\tilde{\psi}}^{2}-\tilde{\mu}\right)\tilde{\psi}=0,
\label{eq5}
\end{equation}
\begin{equation}
\Delta_{\tilde{r}}\tilde{V}_{{\rm {grav}}}={\tilde{\psi}}^{2}.
\label{eq6}
\end{equation} 
The normalization condition $\int \psi^2 d^3r=N$ (where $N$ is the total particle number in the galaxy) gives for the dimensionless chemical potential and total energy
\begin{equation}
\tilde{\mu} =
\frac{\hbar^{2}}{2NGm^{3}R_{c}}
\int_{0}^{\infty}\tilde{r}^{2}d\tilde{r}\left[(\nabla_{\tilde{r}}\tilde{\psi})^{2}+ {\tilde{V}}_{{\rm{grav}}}\tilde{\psi}^{2}+ \tilde{a}\tilde{\psi}^{4} \right],
\label{eq7}
\end{equation}

\begin{equation}
\tilde{E}=\int_{0}^{\infty}\tilde{r}^{2}d\tilde{r}\left[
({\nabla_{\tilde{r}}}\tilde{\psi})^{2}+ \frac{1}{2}{\tilde{V}}_{{\rm{grav}}}\tilde{\psi}^{2}+\frac{1}{2}\tilde{a} \tilde{\psi}^{4}
\right].
\label{eq8}
\end{equation}
The total energy is given by $E=\left(\hbar^{4}/4G m^{4}R_{c}^{3}\right)\tilde{E}$. The dimensionless mass of the BEC halo at position $\tilde{r}$ is also given by
\begin{equation}
\tilde{M}(\tilde{r})=\int_{0}^{\tilde{r}}\tilde{r'}^2 d\tilde{r'}\tilde{\psi}(\tilde{r'})^{2},
\label{eq9}
\end{equation}
and is related to the the mass profile by: 
$M(r)=\left(\hbar^{2}/2G m^{2}R_{c}\right)\tilde{M}(\tilde{r})$. The mass of the dark matter core, $M(R_{c})=M$, is listed in table \ref{table1}. The central density is given by $\rho_c=\chg{m}\,\psi^2(0)$. Finally, the tangential velocity of the fractional BEC dark matter is\cite{Cd12}:
\begin{equation}
V_{\rm{tg}}(r)=\sqrt{\frac{GM(r)}{r}}.
\label{eq10}
\end{equation}

\subsection{Boundary Conditions}

We numerically solve the GPP equations (\ref{eq5}-\ref{eq6}) in the domain $ \tilde{r} \in [0, \tilde{r}_{max}=10]$, to ensure the grid is sufficiently large to capture all relevant physical behavior. The boundary conditions are defined as follows:
\subsubsection*{Inner Boundary Conditions}
To ensure the physical regularity of our numerical solution at the galactic center, we enforce the vanishing of the first derivatives for both the condensate wave function and the gravitational potential:

\begin{equation}
\begin{cases}
\frac{\partial}{\partial\tilde{r}}\tilde{\psi}|_{\tilde{r}=0}&=0,\\ 
\frac{\partial}{\partial\tilde{r}}\tilde{V}_{\rm {grav}}|_{\tilde{r}=0}&=0.
\end{cases}
\label{eq11}
\end{equation}
As established in recent GPP numerical studies \cite{Cd29}, these boundary conditions guarantee finite and smooth solutions at the center and a localized core with vanishing density at infinity.

\subsubsection*{Outer Boundary Conditions}

The conditions at $\tilde{r}_{max}$ must reflect the asymptotic behavior of the halo. We first impose that the condensate density to decay to zero, i.e. $\tilde{\psi} (\tilde{r}_{max})= 0$. Second, the boundary condition for the gravitational potential must be set taking into account the global mass distribution. 

\begin{equation}
\tilde{V}_{{\rm {grav}}}(\tilde{r}_{max}) = -\frac{\tilde{M}(\tilde{r}=1)}{\tilde{r}_{max}}.
\label{eq12}
\end{equation}
This boundary condition puts the quantum core inside the larger dark matter halo. The SPARC observations show rotation curves that can only be correctly reproduced using the total halo potential.

Lastly, we verify our model by examining its internal consistency. We check that the numerical gravitational force at the boundary is equal to the core mass, equation (\ref{eq9}), and that the numerical derivative of the potential at the boundary satisfies perfectly:
\begin{equation}
    \left. \frac{\partial\tilde{V}_{{\rm {grav}}}}{\partial\tilde{r}} \right|_{\tilde{r}=\tilde{r}_{max}} =  \frac{\tilde{M}(\tilde{r}=1)}{\tilde{r}^{2}_{max}}.
\end{equation}
This will ensure that our method correctly captures the internal gravity of the BEC core while maintaining consistency with the global asymptotic boundary conditions of the entire galactic halo.

The way in which the boundary conditions are treated for the GPP equations is a critical factor in governing the stability of the solutions. Notably, Guzm\'an et al.~\citep{Cd20} used a much different set of boundary conditions. They described the halos's central density using a Thomas--Fermi profile, with its extent set by the galaxy's radius. The gravitational potential satisfied monopolar boundary conditions, which guaranteed that it was finite at the center and tended to zero at large distances. This choice allowed them to relate directly the galaxy radius, the scattering length, and the mass of the condensate particles. We suggest that the particular choice of boundary conditions may account in part for the strong constraints that they derived for the B\"ohmer--Harko model. In contrast, Ure\~na-L\'opez et al.~\citep{Cd37} imposed boundary conditions at infinity similar to those used in the present work, and determined the central boundary conditions through a shooting method. We believe that this approach can lead to overshooting, unless the numerical process is carefully controlled, for example by using an appropriate relaxation scheme.

The numerical algorithm proceeds iteratively as follows: starting with an initial guess for the chemical potential (\ref{eq7}), we solve the GPP equations (\ref{eq5}-\ref{eq6}). The output of this process is then substituted into (\ref{eq7}) to obtain an updated value for $\tilde{\mu}$. This process continues until the difference between two successive values of $\tilde{\mu}$ is less than a predefined tolerance ($\sim 10^{-6}$). The tangential velocity is finally calculated using (\ref{eq10}). 
The parameter $N$ is determined self-consistently for each galaxy using its mass profile $M(r)$ (Equation \ref{eq9}).

But the algorithm still needs two inputs: the mass of the unknown particles and their scattering length. We invert the fractional BEC scaling relations to ensure that these parameters are physically justified, aligning them with the structural parameters observed in our SPARC sample. 
Recent work has employed the internal kinematics of dark-matter-dominated SPARC galaxies to limit the properties of BEC dark matter, finding that these systems possess core masses of roughly $10^8 M_{\odot}$ and core radii of order kiloparsecs \cite{Cd32}. 

We obtain core radii $Rc\approx2-8\,{\rm kpc}$ and core masses $M\approx \,10^{9}M_{\bigodot}$ for galaxies in our sample. In the fractional BEC picture these scales constrain the particle mass and scattering length to ($m\approx\,10^{-26}\text{eV}/c^2$, $a\approx\,10^{-82}\text{cm}$) obtained by inverting the BEC scaling relations so as to reproduce the observed inner rotation curves {\it and} satisfy the condition $R_c>R_{\rm disk}$, where $R_{\rm disk}$ is the radius of the stellar disk (taken from \cite{Cd27}). \chg{These values should be understood as a representative parameter scale. The sensitivity test of Section~\ref{sec:sens} (Table~\ref{tab:sidebyside}) illustrates how the mass scale of the solutions changes with $m$, but since it relies on the constraint $R_c>R_{\rm disk}$ and on a fixed dimensionless solution, we do not regard it as a determination of $m$ and $a$.}

The very small inferred scattering length, $a\approx\,10^{-82}\text{cm}$, seems broadly consistent with estimates obtained within the string axiverse scenario \cite{Cd24}, where compactification of extra dimensions is expected to produce a huge variety of ultralight axion-like particles. In this context, the scattering length is $a\approx m/(8\pi \gamma_a^2)$ and originates from a $\phi^4$ self-interaction of the axion field with coupling $\lambda\sim m^2/\gamma_a^2$, where $\gamma_a$ is the axion decay constant. Using a grand-unification-scale decay constant $\gamma_a \simeq 10^{16}$ GeV, and an ultralight axion mass $m\simeq 10^{-26}$ eV$/c^2$, gives $a \simeq 10^{-82}$ cm, which is reasonably close to our fitted value. This correspondence should be regarded as suggestive rather than definitive, but it does raise the possibility that the string axiverse scenario could offer a top-down theoretical rationale for the parameter regime explored in this study. \chg{We note that this correspondence concerns only the order of magnitude of $|a|$: for the simplest cosine potential of an axion the quartic self-interaction is attractive, whereas a repulsive interaction ($a>0$) is assumed here, so the identification with a specific axion model remains an open question.}

The effectiveness of the model is tested by presenting the numerical results of the rotation curves of 30 galaxies selected from the sample data obtained by the Spitzer Photometry and Accurate Rotation Curves (SPARC) \cite{Cd27, Cd28}.


\chg{The fits determine the core radius $R_c$ and, through Eq.~(\ref{eq9}), the core mass $M=M(R_c)$ of the condensate (Table~\ref{table1}). For UGC02023, for instance, one finds $M\approx 0.34\times10^{10}\,M_\odot$, which is about $4\%$ of the total SPARC halo mass $M_{\rm DM}=10^{10.89}\,M_\odot$ quoted for this galaxy. This ratio is purely illustrative: it compares the condensate mass inside $R_c$ with the total halo mass of a single galaxy, it is not an output of the fit, and it must not be interpreted as the cosmological fraction of the ultralight component. A consistent definition of that fraction requires a model of the CDM envelope and is left for future work (Section~\ref{sec:density}).}

\subsection{Fitting the SPARC Rotation Curves}

In order to properly model the galactic rotation curves, we need to include the effects of baryons (stars and gas) in addition to the dark matter. The total tangential velocity of the galaxy is obtained by adding up all the components in quadrature, following the approach described in \cite{Cd27}:
\begin{equation}
V_{\mathrm{tot}}(r) =
\sqrt{
V_{\mathrm{tg}}^{2} +
\Upsilon_{\mathrm{disk}}\,\times v_{\mathrm{disk}}|v_{\mathrm{disk}}| +
\Upsilon_{\mathrm{bulge}}\,\times v_{\mathrm{bulge}}|v_{\mathrm{bulge}}| +
v_{\mathrm{gas}}|v_{\mathrm{gas}}|
}
\label{vtot}
\end{equation}
where $V_{\mathrm{tg}}$ represents the contribution from the fractional Bose-Einstein condensate dark matter (Eq. \ref{eq10}), $v_{\mathrm{gas}}$ represents the contribution from the gaseous component, $v_{\mathrm{disk}}$ and $v_{\mathrm{bulge}}$ represent the contribution from the stellar disk and bulge. \chg{Here $\Upsilon_{\rm disk}$ and $\Upsilon_{\rm bulge}$ are the stellar mass-to-light ratios of the disk and of the bulge (in solar units at $3.6\,\mu$m), by which the SPARC velocity contributions $v_{\rm disk}$ and $v_{\rm bulge}$, computed for unit mass-to-light ratio, are scaled; the fiducial SPARC values are $\Upsilon_{\rm disk}=0.5$ and $\Upsilon_{\rm bulge}=0.7$~\cite{Cd27}.} The absolute values are included to cover regions where the gas contribution may formally be negative, because of central depressions in the neutral atomic hydrogen (H\,\textsc{i}) distribution. 
We adopt the usual treatment of the baryonic components as fixed according to observations, as done in the SPARC database \cite{Cd27,Cd28}. However, because the galaxies in our sample are morphologically classified as bulgeless, the bulge contribution is set to zero in the rotation curve decomposition. Thus, $\Upsilon_{\mathrm{disk}}$ is the only free parameter for the stellar component. The best-fit values of $\Upsilon_{\mathrm{disk}}$ obtained for each galaxy are listed in table \ref{table1}. 
The goodness-of-fit for each galaxy is established by minimizing the reduced chi-squared statistic (\(\chi ^{2}\)), which is defined as:
\begin{equation}
    \chi^2 = \frac{1}{N_d - N_f} \sum_{i=1}^{N} \left( \frac{V_{\mathrm{obs},i} - V_{\mathrm{tot},i}}{{\sigma_{i}}} \right)^2 .
    \label{eqchisq}
\end{equation}
In this expression, $V_{\mathrm{obs},i}$ represents the observed rotation velocity, $V_{\mathrm{tot},i}$ is the velocity expected from our model (given by Eq.~\ref{vtot}), and $\sigma_i$ is the measurement error at the $i$-th radial point from \cite{Cd27}. $N_d$ is the total number of data points for a given galaxy, and $N_f = 2$ is the number of free parameters in the fit, $\Upsilon_{\mathrm{disk}}$ and $R_c$. We find these parameters by solving the full GPP system over the parameter space $(R_c, \Upsilon_{\mathrm{disk}})$, and selecting the combination which yields the minimum $\chi^2$ with respect to the observed SPARC rotation curves. 
The best-fit parameters and their associated \(\chi ^{2}\) values are shown in table~\ref{table1}. The extracted values show that the radius of the BEC core is always larger than \(R_{\mathrm{disk}}\)\chg{, which is imposed as a constraint in the fits. In the fitted radial range the condensate supplies the gravitational potential that is required in addition to the baryonic one.}\\
\newpage
\begin{center}
\captionof{table}{Fitting parameters $R_{c}$ and $\Upsilon_{\mathrm{disk}}$ with $m=1.404\times 10^{-26}$ eV$/c^2$ and $a=1.1\times 10^{-82}$ cm. $R_{\text{disk}}$ is taken from  \cite{Cd27}. $M=M(R_c)$ is computed from Eq. (\ref{eq9}). }
\begin{tabular}{lccccc}
\toprule
 \textbf{Galaxy} &$R_{\text{disk}}({\rm kpc})$ & \textbf{$R_{c}({\rm kpc})$} & \textbf{$M (10^{10}M_{\odot})$}  & \textbf{$\Upsilon_{\mathrm{disk}}$}& \textbf{$\chi^{2}$}\\
\midrule
      D564-8   & 0.610 & 5.804 & 0.252 & 0.729 & 0.580 \\
      DDO064   & 0.690 & 3.930 & 0.366 & 1.500 & 0.550 \\
      KK98-251 & 1.340 & 5.297 & 0.275 & 1.500 & 0.460 \\
      NGC0100  & 1.660 & 5.195 & 0.281 & 1.105 & 0.678 \\
      NGC2976  & 1.010 & 3.089 & 0.458 & 0.877 & 0.340 \\
      NGC3521  & 2.400 & 4.324 & 0.335 & 0.590 & 0.180 \\
      NGC3949  & 3.590 & 3.867 & 0.372 & 0.528 & 0.334 \\
      NGC3953  & 4.890 & 7.791 & 0.189 & 0.820 & 0.607 \\      
      NGC4068  & 0.590 & 4.119 & 0.351 & 0.482 & 0.178 \\
      NGC4088  & 2.580 & 5.739 & 0.255 & 0.431 & 0.499 \\
      NGC4183  & 2.790 & 7.027 & 0.209 & 1.500 & 0.683 \\
      NGC4389  & 2.790 & 3.616 & 0.396 & 0.100 & 0.321 \\
      UGC01281 & 1.630 & 4.653 & 0.312 & 1.500 & 0.727 \\
      UGC02023 & 1.550 & 4.244 & 0.341 & 0.261 & 0.010 \\  
      UGC02455 & 0.990 & 5.030 & 0.290 & 0.100 & 0.564 \\
      UGC04278 & 2.210 & 4.445 & 0.326 & 1.235 & 0.397 \\
      UGC04483 & 0.180 & 4.091 & 0.353 & 1.241 & 0.857 \\
      UGC05005 & 3.200 & 7.494 & 0.196 & 1.366 & 0.259 \\
      UGC05414 & 1.470 & 4.743 & 0.306 & 1.067 & 0.377 \\
      UGC05918 & 1.660 & 8.900 & 0.165 & 1.036 & 0.172 \\
      UGC07089 & 2.260 & 5.993 & 0.244 & 0.905 & 0.131 \\
      UGC07232 & 0.290 & 2.252 & 0.582 & 0.443 & 0.211 \\
      UGC07261 & 1.200 & 5.328 & 0.274 & 1.500 & 0.928 \\
      UGC07323 & 2.260 & 4.954 & 0.294 & 0.897 & 0.253 \\
      UGC07559 & 0.580 & 5.010 & 0.291 & 0.922 & 0.283 \\
      UGC07577 & 0.900 & 6.246 & 0.235 & 0.648 & 0.125 \\   
      UGC07866 & 0.610 & 5.285 & 0.276 & 1.500 & 0.035 \\
      UGC08837 & 1.720 & 5.019 & 0.290 & 0.420 & 0.325 \\
      UGCA281  & 1.720 & 3.828 & 0.376 & 0.918 & 0.236 \\
      UGCA444  & 0.830 & 4.073 & 0.354 & 1.500 & 0.922 \\ 
    
\bottomrule
\end{tabular}
\label{table1}
\end{center}

\subsection{Density model and scope of the fits}
\label{sec:density}
\chg{The scenario considered here is that of a fractional condensate: a core of ultralight bosons embedded in an extended halo of cold dark matter~\cite{Cd25,Cd26}. The dark-matter density is therefore the sum of two components, $\rho_{\rm DM}(r)=\rho_{\rm BEC}(r)+\rho_{\rm CDM}(r)$, where $\rho_{\rm BEC}=m\psi^2$ is the stationary solution of Sections 2.1--2.3, which vanishes at the outer boundary of the computational domain by construction ($\tilde\psi(\tilde r_{\max})=0$), and $\rho_{\rm CDM}$ describes the cold component, which dominates the outskirts. In the fits of Section~\ref{sec:analysis} only the condensate is included in the total velocity, Eq.~(\ref{vtot}): no CDM envelope is modeled and the CDM contribution is neglected. The fits therefore test whether a condensate with parameters $(R_c,\Upsilon_{\rm disk})$ is compatible with the inner kinematics of each galaxy, and we make no statement about the profile or the mass of the outer halo.

We stress that neglecting the CDM component is not obviously justified at the level of the enclosed mass. As an illustration, for a halo of mass $M_{200}=10^{10.89}\,M_\odot$ (the value quoted above for UGC02023) and an assumed concentration $c\simeq12$, an NFW profile encloses of order $5\times10^{9}\,M_\odot$ within $5$ kpc, which is comparable to the condensate masses of Table~\ref{table1}. A fit including both components, and a consistent definition of the fraction of dark matter in the condensate, are therefore left for future work.

The relation to the core--halo configurations found in simulations of ultralight dark matter should also be stated clearly. In those simulations, which do not include self-interaction (Refs.~\cite{Cd33,Bar18,Liu23}), a soliton core whose size is of the order of the de Broglie wavelength at the halo centre is surrounded by an NFW-like envelope, with a transition at a radius of a few soliton radii and a soliton--halo mass relation $M_c\propto M_h^{1/3}$~\cite{Cd33}. These results assume that quantum pressure is the only support against gravity. For $m=1.4\times10^{-26}$ eV$/c^2$ the de Broglie wavelength $\lambda_{\rm dB}=2\pi\hbar/(mv)$ is about $9$ Mpc for $v=100$ km s$^{-1}$ and about $30$ Mpc for $v=30$ km s$^{-1}$, much larger than any galaxy, and the interaction parameter is $\tilde a\sim10^{2}$--$10^{3}$ for the values of $R_c$ in Table~\ref{table1}, so that the self-interaction is not negligible with respect to gravity. The simulation-based relations therefore cannot be assumed to hold in the regime studied here.}

\subsection{Role of the baryons in the condensate equation}
\label{sec:baryons}
\chg{In Eqs.~(\ref{eq3})--(\ref{eq4}) only the self-gravity of the condensate is included, while the baryons enter the fits through the total velocity, Eq.~(\ref{vtot}). The baryons do act on the condensate through their gravitational potential, and for heavier bosons this effect can be as important as their direct contribution to the velocity~\cite{Kor26}. We therefore estimate its size in our case. Writing the baryonic potential energy per boson as $m\Phi_{\rm bar}$, with $\Phi_{\rm bar}\sim v_{\rm bar}^2$ (from $d\Phi_{\rm bar}/dr=v_{\rm bar}^2/r$), the corresponding term in the dimensionless Gross--Pitaevskii equation (\ref{eq5}) is, up to factors of order unity,
\begin{equation}
\tilde V_{\rm bar}=\frac{2m^{2}R_c^{2}}{\hbar^{2}}\,\Phi_{\rm bar}.
\label{eqvbar}
\end{equation}
This must be compared with the kinetic term, of order $\tilde r_{\max}^{-2}\sim10^{-2}$. Table~\ref{tab:vbar} gives $\tilde V_{\rm bar}$ for $m=1.404\times10^{-26}$ eV$/c^2$ and for typical core radii and baryonic velocities of the sample.}
\begin{center}
\captionof{table}{\chg{Order of magnitude of $\tilde V_{\rm bar}$, Eq.~(\ref{eqvbar}), for $m=1.404\times10^{-26}$ eV$/c^2$.}}
\begin{tabular}{lccc}
\toprule
$R_c$ (kpc) & $v_{\rm bar}=20$ km/s & $50$ km/s & $80$ km/s\\
\midrule
2.25 & $2\times10^{-7}$ & $1\times10^{-6}$ & $3\times10^{-6}$\\
4    & $7\times10^{-7}$ & $4\times10^{-6}$ & $1\times10^{-5}$\\
8.9  & $3\times10^{-6}$ & $2\times10^{-5}$ & $5\times10^{-5}$\\
\bottomrule
\end{tabular}
\label{tab:vbar}
\end{center}
\chg{For these parameters $\tilde V_{\rm bar}$ is between two and five orders of magnitude smaller than the kinetic term. Treating it as a perturbation of Eq.~(\ref{eq5}), the relative change of the condensate profile is of the same order as this ratio, so that the backreaction of the baryons on the condensate is negligible and it is consistent to include the baryons only through Eq.~(\ref{vtot}). 

The conclusion depends strongly on the boson mass, since $\tilde V_{\rm bar}\propto m^{2}$: for $m\sim10^{-22}$ eV$/c^2$ and $R_c\sim1$ kpc one finds $\tilde V_{\rm bar}\sim10$, and the baryons strongly modify the condensate, in agreement with Ref.~\cite{Kor26}. Our treatment is justified only for the very small mass considered here and should not be extrapolated to heavier bosons.}

\subsection{Cosmological and dynamical constraints}
\label{sec:constraints}
\chg{A scalar field of mass $m=1.4\times10^{-26}$ eV$/c^2$ starts to oscillate when $H\simeq mc^2/\hbar\simeq2\times10^{-11}$ s$^{-1}$, which for a standard expansion history corresponds to a redshift $z\sim3\times10^{4}$, well before matter--radiation equality, and it behaves as a matter component afterwards. A component of this kind is constrained by CMB and large-scale-structure data~\cite{Hlo15}, and these constraints are the reason why we consider only a sub-dominant (fractional) condensate. We do not claim that the fractional assumption removes them: the allowed fraction at this mass has not been evaluated here, and the illustrative ratio given above for a single galaxy cannot be used for that purpose.

Dynamical constraints from the heating of stellar systems~\cite{Dal22} arise from the granular fluctuations of the ultralight field, whose characteristic size is the de Broglie wavelength. These bounds were derived for masses for which this wavelength is comparable to the size of the stellar systems considered. For $m=1.4\times10^{-26}$ eV$/c^2$ the wavelength is of order $10$ Mpc or larger for velocities of order $100$ km s$^{-1}$ or below, much larger than any galaxy, so that the field is coherent over the whole halo and no granules exist on galactic scales. The mechanism of Ref.~\cite{Dal22} therefore does not apply directly to our regime, and a dedicated analysis of the response of stellar orbits to a coherent, time-dependent field would be needed. We regard the scenario with $m\sim10^{-26}$ eV$/c^2$ as phenomenologically motivated by the rotation-curve fits; a complete cosmological analysis is beyond the scope of this paper.}

\section{Numerical Solutions and Astrophysical Validation}
\label{sec:analysis}
We now use our fractional BEC dark matter model to the inner kinematics of the SPARC database. 
Beyond the local Thomas–Fermi approximation, the quantum pressure term provides an important dispersive contribution, regularizing the central density profile. We emphasize that the central density \(\rho _{c}\) is not a freely adjustable parameter but is instead derived from the GPP equations.

\subsection{Rotation Curve Analysis}

\begin{figure*}[h]
    \centering
      \begin{subfigure}{0.22\textwidth}
        \centering 
        \textbf{D564-8}\par 
        \vspace{0.02cm} 
        \includegraphics[width=\textwidth]{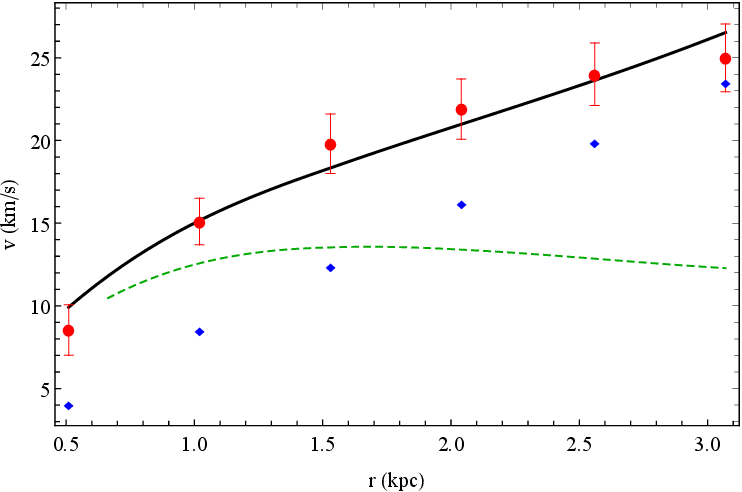}
    \end{subfigure}
    \hfill
    \begin{subfigure}{0.22\textwidth}
        \centering 
        \textbf{DDO064}\par 
        \vspace{0.02cm} 
        \includegraphics[width=\textwidth]{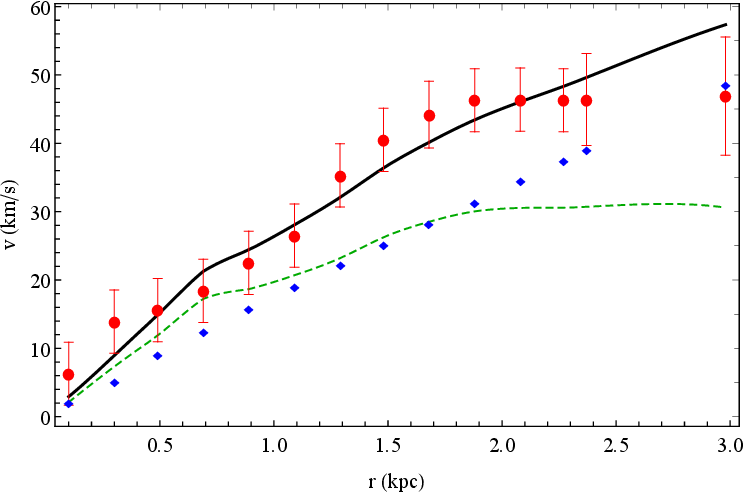}
    \end{subfigure}
    \hfill
    \begin{subfigure}{0.22\textwidth}
        \centering 
        \textbf{KK98-251}\par 
        \vspace{0.02cm} 
        \includegraphics[width=\textwidth]{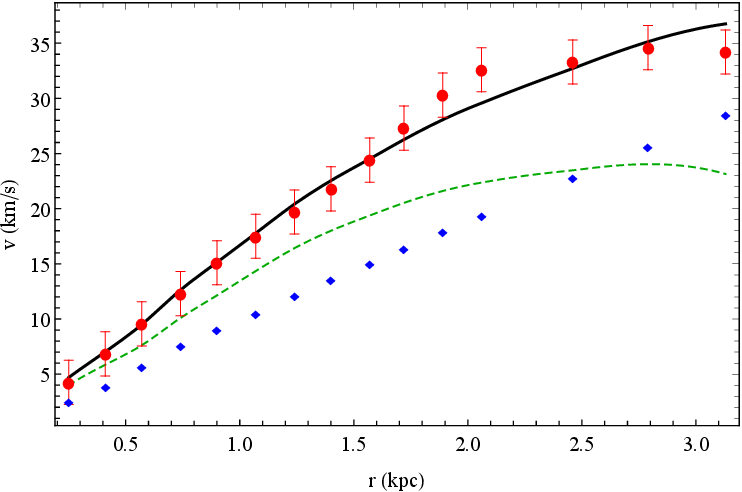}
    \end{subfigure}
    \hfill
    \begin{subfigure}{0.22\textwidth}
        \centering 
        \textbf{NGC0100}\par 
        \vspace{0.02cm} 
        \includegraphics[width=\textwidth]{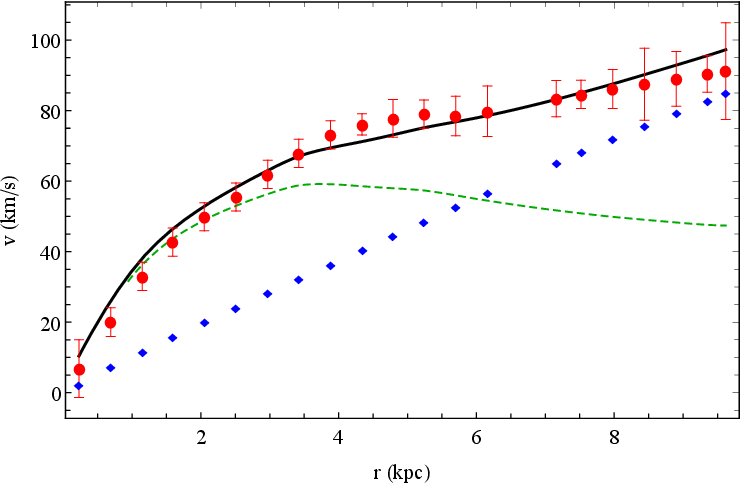}
    \end{subfigure}
        \vspace{0.02cm} 
    \begin{subfigure}{0.22\textwidth}
        \centering \textbf{NGC2976} \\
        \includegraphics[width=\textwidth]{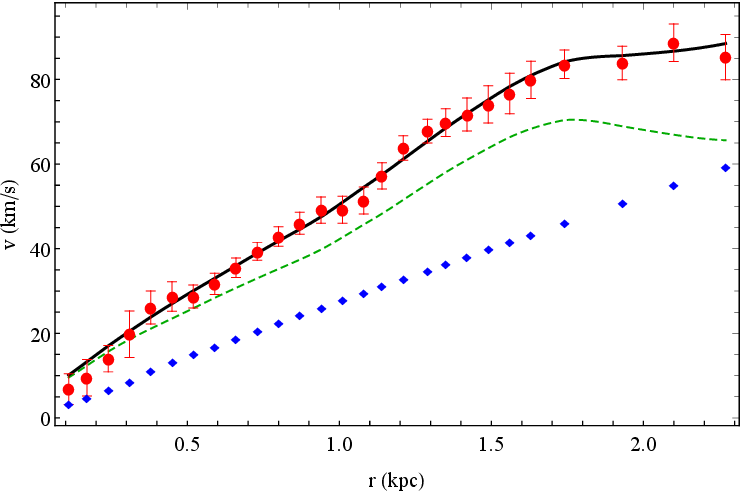}
    \end{subfigure}
        \hfill
    \begin{subfigure}{0.22\textwidth}
        \centering \textbf{NGC3521} \\
        \includegraphics[width=\textwidth]{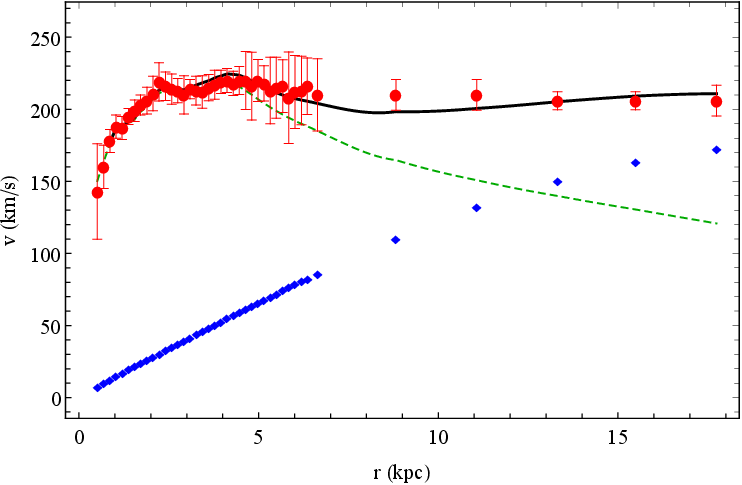}
    \end{subfigure}
    \hfill
    \begin{subfigure}{0.22\textwidth}
        \centering \textbf{NGC3949} \\
        \includegraphics[width=\textwidth]{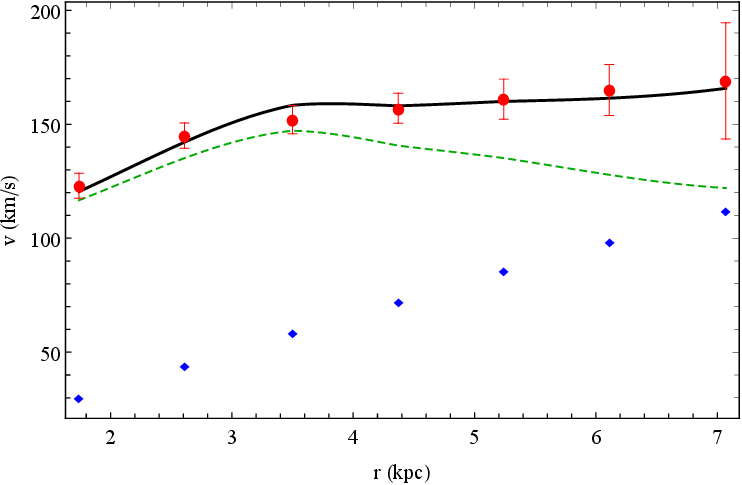}
    \end{subfigure}
    \hfill
       \begin{subfigure}{0.22\textwidth}
        \centering \textbf{NGC3953} \\
        \includegraphics[width=\textwidth]{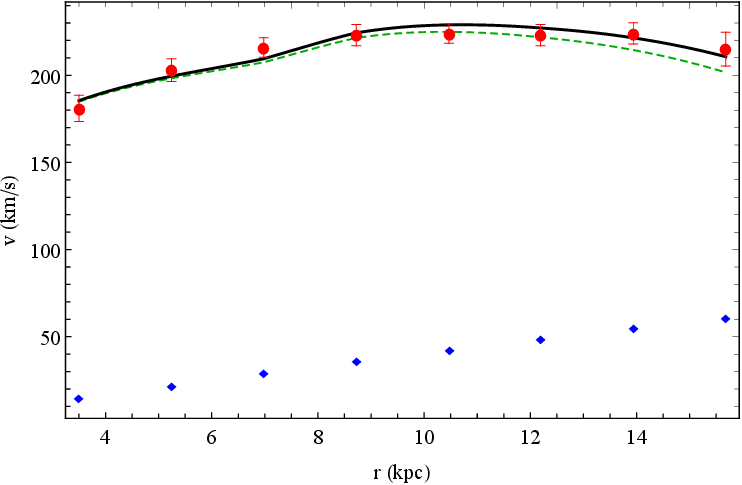}
    \end{subfigure}
        \vspace{0.1cm}
       \begin{subfigure}{0.22\textwidth}
        \centering \textbf{NGC4068} \\
        \includegraphics[width=\textwidth]{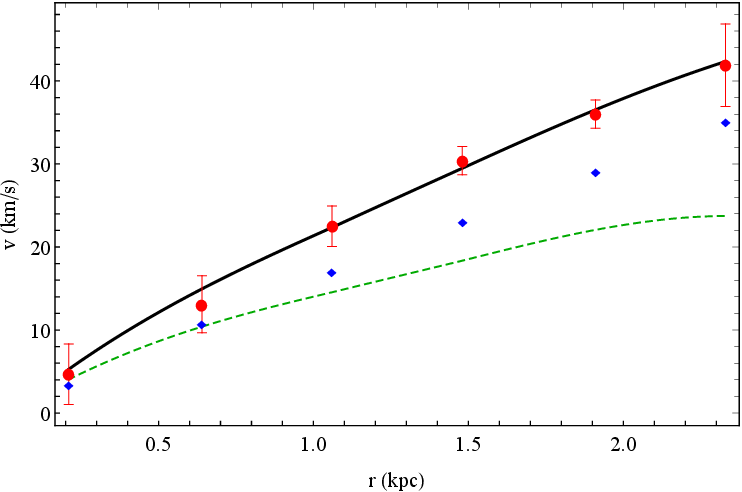}
    \end{subfigure}
    \hfill
    \begin{subfigure}{0.22\textwidth}
        \centering \textbf{NGC4088} \\
        \includegraphics[width=\textwidth]{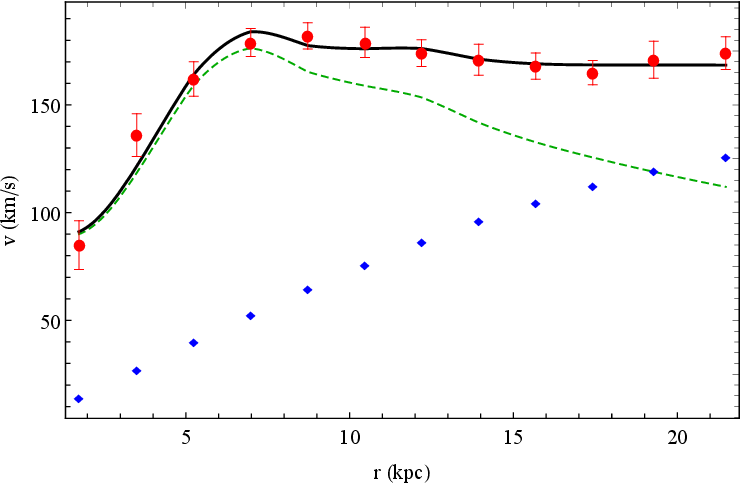}
    \end{subfigure}
    \hfill
    \begin{subfigure}{0.22\textwidth}
        \centering \textbf{NGC4183} \\
        \includegraphics[width=\textwidth]{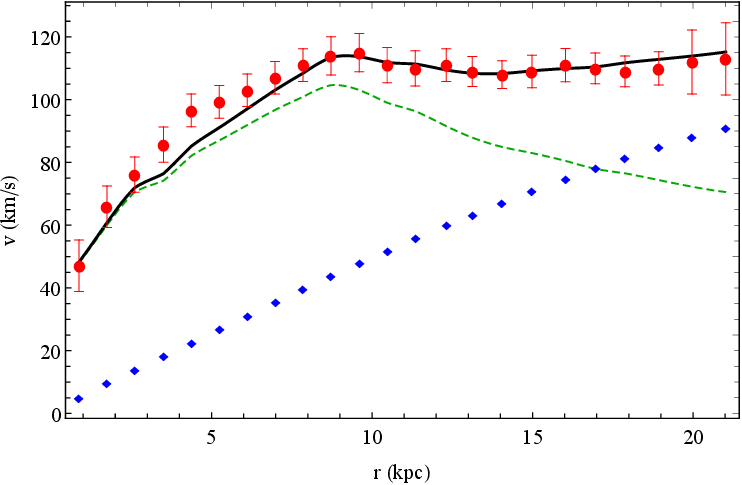}
    \end{subfigure}
    \hfill
       \begin{subfigure}{0.22\textwidth}
        \centering \textbf{NGC4389} \\
        \includegraphics[width=\textwidth]{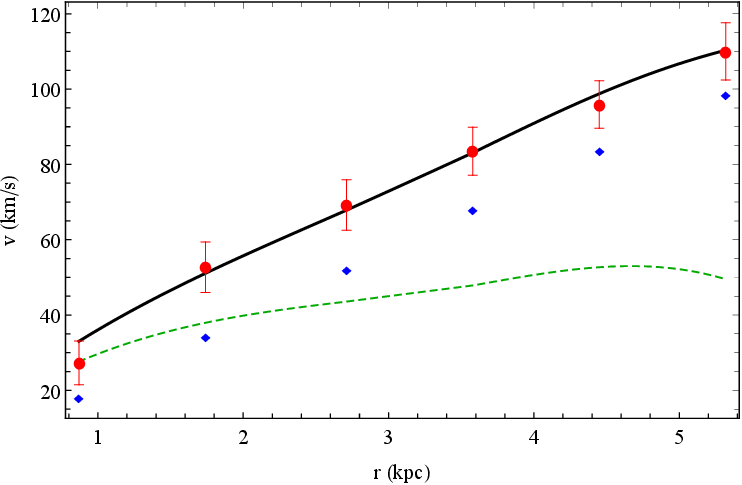}
    \end{subfigure}
    \vspace{0.1cm}
        \begin{subfigure}{0.22\textwidth}
        \centering \textbf{UGC01281} \\
        \includegraphics[width=\textwidth]{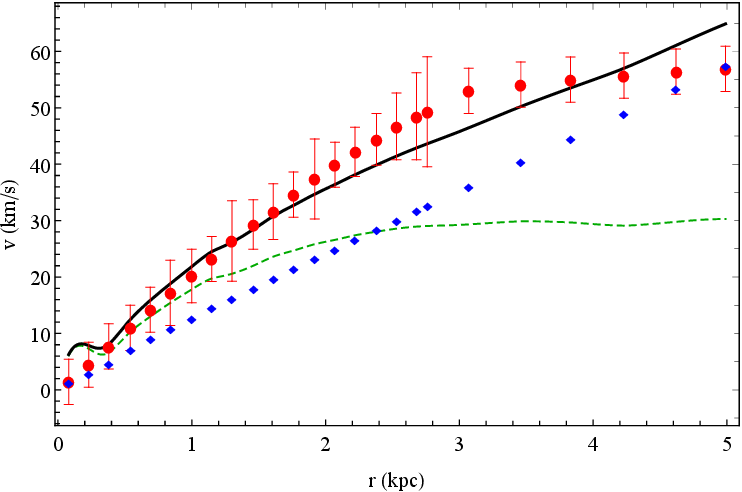}
    \end{subfigure}
        \hfill
    \begin{subfigure}{0.22\textwidth}
        \centering \textbf{UGC02023} \\
        \includegraphics[width=\textwidth]{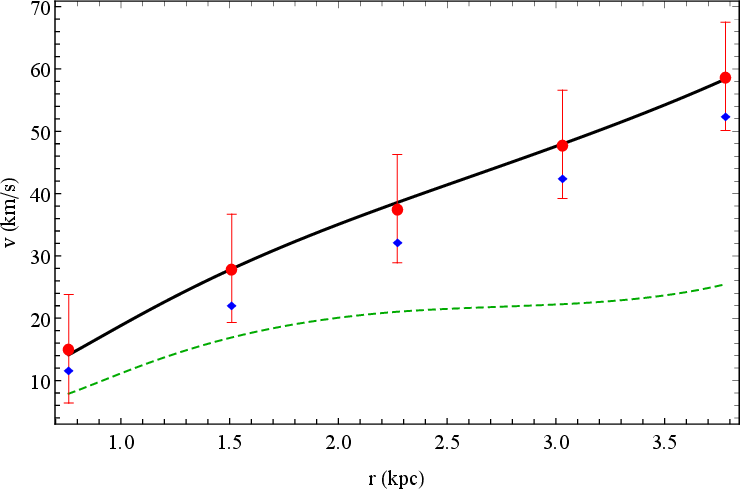}
    \end{subfigure}
        \hfill
        \begin{subfigure}{0.22\textwidth}
        \centering 
        \textbf{UGC02455}\par 
        \includegraphics[width=\textwidth]{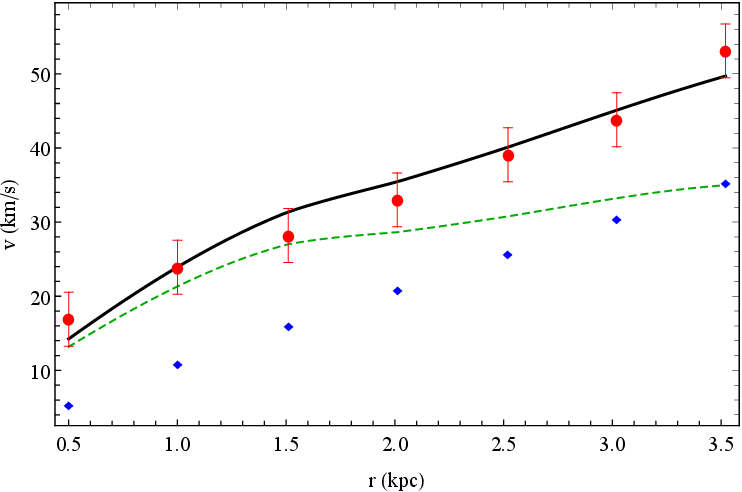}
    \end{subfigure}
        \hfill
        \begin{subfigure}{0.22\textwidth}
        \centering 
        \textbf{UGC04278}\par 
        \includegraphics[width=\textwidth]{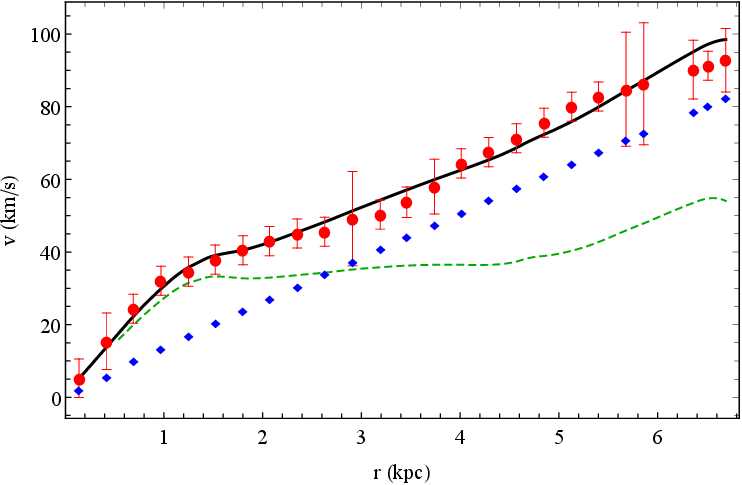}
    \end{subfigure}
 \caption{
 Tangential velocity in the fractional BEC dark matter model (with particle mass $m=1.404\times 10^{-26}$ eV$/c^2$ and scattering length $a=1.1\times 10^{-82}$ cm). The green dashed curve represents the Newtonian contribution from visible baryonic matter alone; the blue diamonds denote the contribution from fractional BEC dark matter alone; and the solid black curve shows the total tangential velocity, combining both baryonic and BEC dark matter contributions.}
    \label{fig1}
\end{figure*}

\begin{figure*}[h]
    \centering
    \begin{subfigure}{0.22\textwidth}
        \centering 
        \textbf{UGC04483}\par 
        \includegraphics[width=\textwidth]{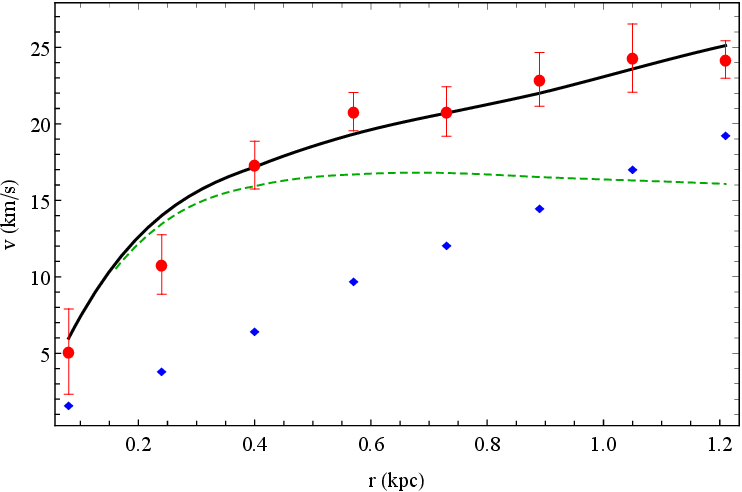}
    \end{subfigure}
    \hfill
    \begin{subfigure}{0.22\textwidth}
        \centering 
        \textbf{UGC05005}\par 
        \includegraphics[width=\textwidth]{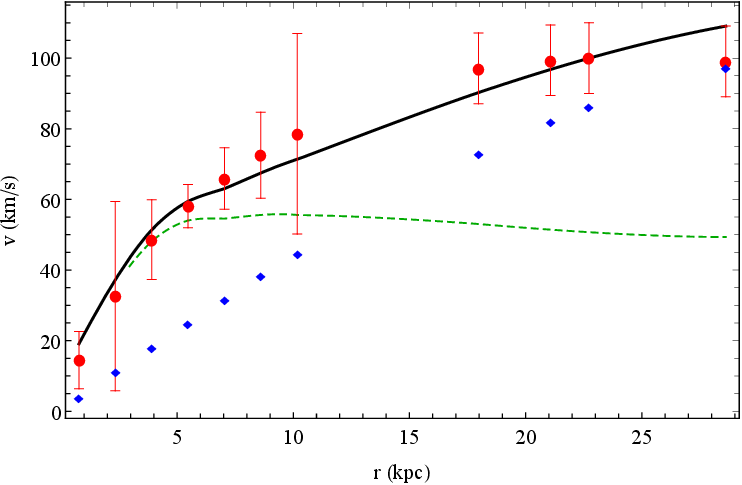}
    \end{subfigure}
        \hfill
    \begin{subfigure}{0.22\textwidth}
        \centering \textbf{UGC05414} \\
        \includegraphics[width=\textwidth]{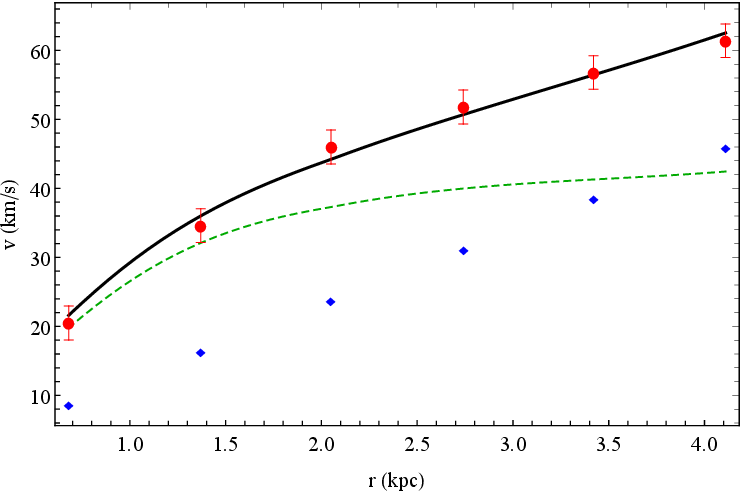}
    \end{subfigure}
        \hfill
    \begin{subfigure}{0.22\textwidth}
        \centering \textbf{UGC05918} \\
        \includegraphics[width=\textwidth]{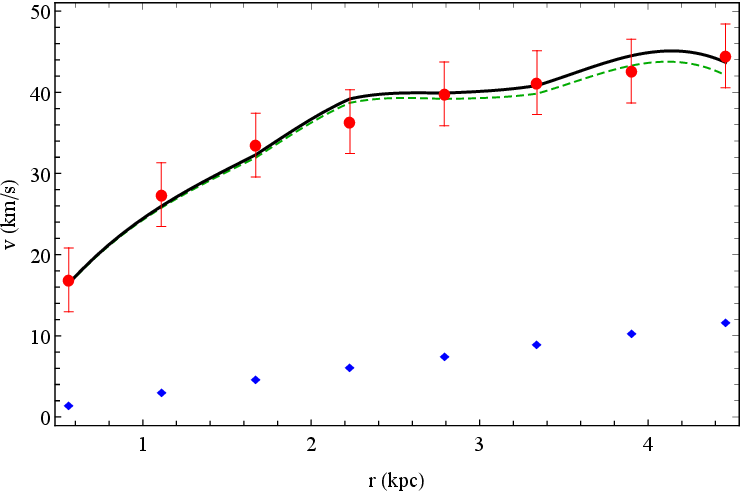}
    \end{subfigure}
            \vspace{0.1cm} 
        \begin{subfigure}{0.22\textwidth}
        \centering \textbf{UGC07089} \\
        \includegraphics[width=\textwidth]{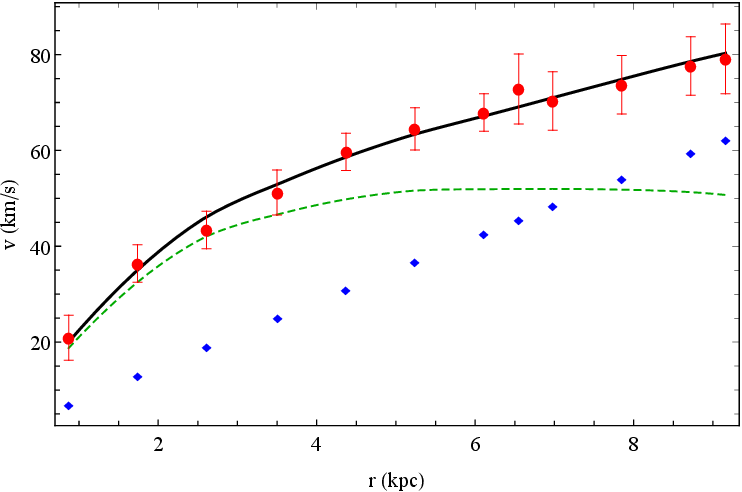}
    \end{subfigure}
     \hfill
     \begin{subfigure}{0.22\textwidth}
        \centering \textbf{UGC07232} \\
        \includegraphics[width=\textwidth]{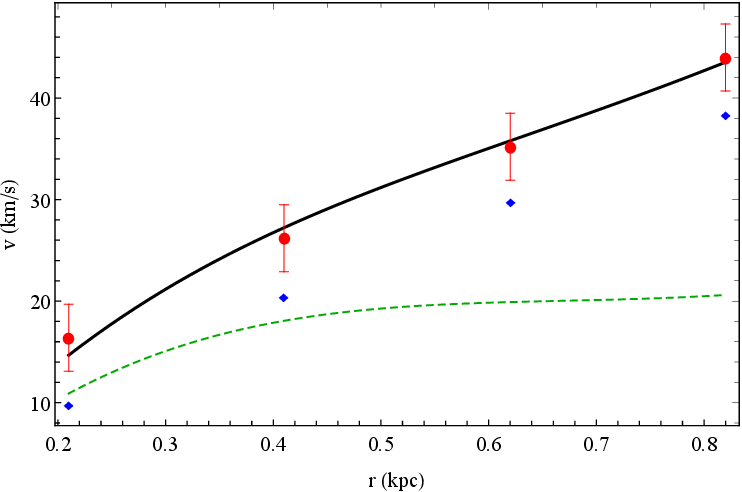}
    \end{subfigure}
        \hfill
    \begin{subfigure}{0.22\textwidth}
        \centering \textbf{UGC07261} \\
        \includegraphics[width=\textwidth]{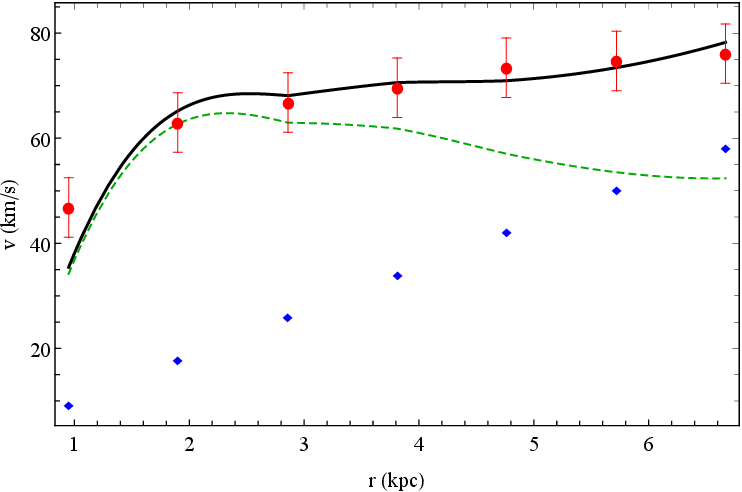}
    \end{subfigure}
        \hfill
    \begin{subfigure}{0.22\textwidth}
        \centering \textbf{UGC07323} \\
        \includegraphics[width=\textwidth]{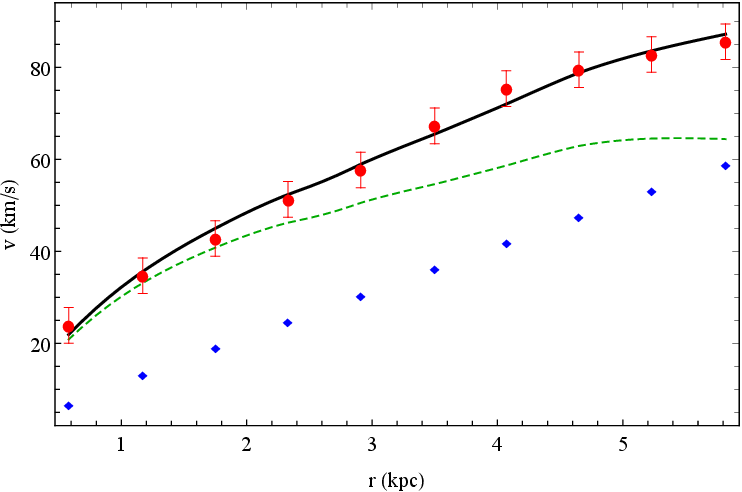}
    \end{subfigure}
     \vspace{0.1cm} 
 \begin{subfigure}{0.22\textwidth}
        \centering \textbf{UGC07559} \\
        \includegraphics[width=\textwidth]{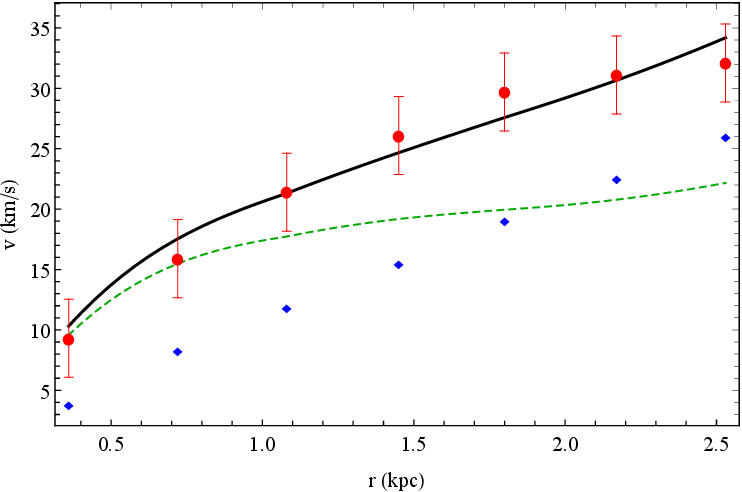}
    \end{subfigure}
    \hfill
    \begin{subfigure}{0.22\textwidth}
        \centering \textbf{UGC07577} \\
        \includegraphics[width=\textwidth]{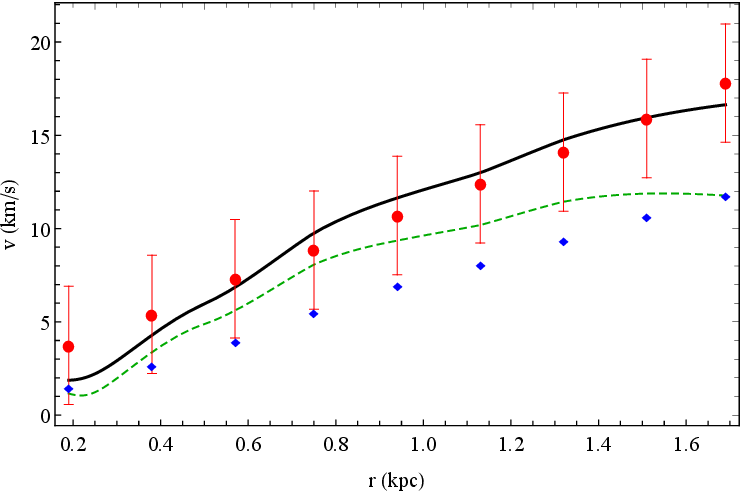}
    \end{subfigure}
      \hfill
    \begin{subfigure}{0.22\textwidth}
        \centering \textbf{UGC07866} \\
        \includegraphics[width=\textwidth]{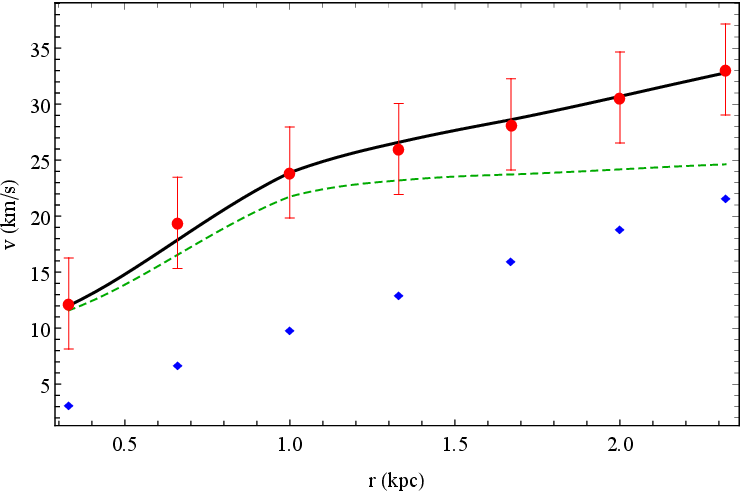}
    \end{subfigure}
      \hfill
            \begin{subfigure}{0.22\textwidth}
        \centering \textbf{UGC08837} \\
        \includegraphics[width=\textwidth]{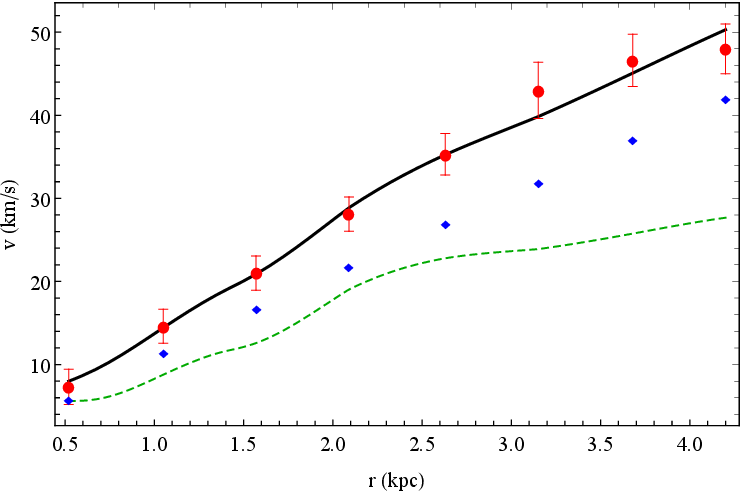}
    \end{subfigure} 
             \vspace{0.1cm}
    \begin{subfigure}{0.22\textwidth}
        \centering \textbf{UGCA281} \\
        \includegraphics[width=\textwidth]{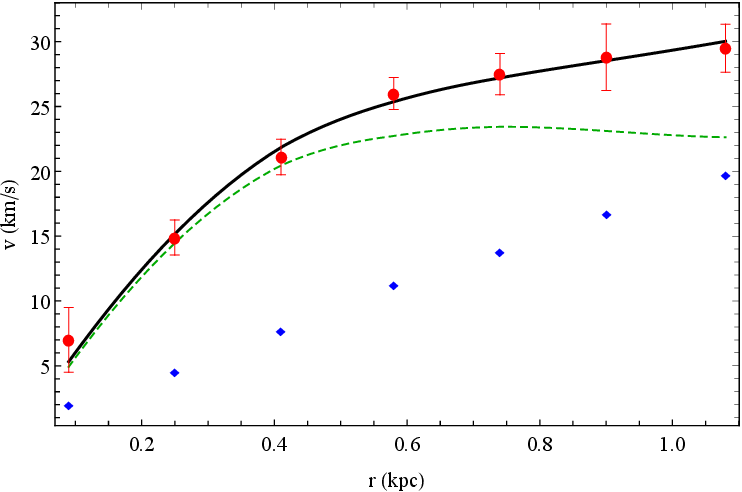}
    \end{subfigure}
    \begin{subfigure}{0.22\textwidth}
        \centering \textbf{UGCA444} \\
        \includegraphics[width=\textwidth]{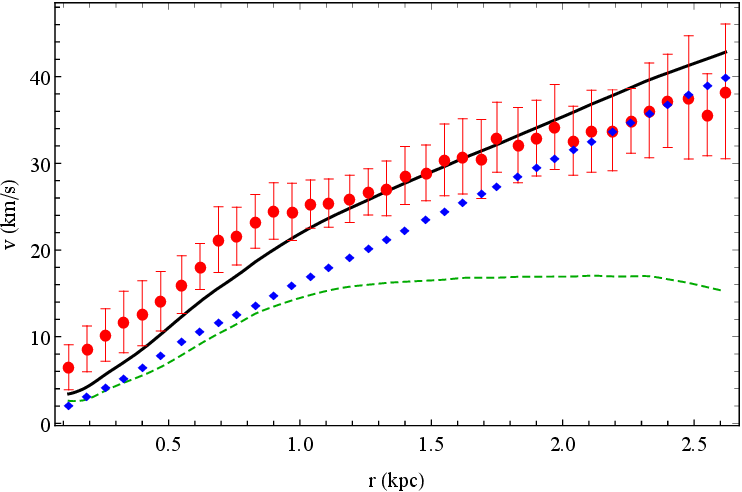}
    \end{subfigure}
 \caption{Same as Fig. \ref{fig1}, \chg{with the same particle mass and scattering length, for the galaxies named in each panel.}}
\label{fig2}
\end{figure*}

The outcomes are illustrated in figures \ref{fig1} and \ref{fig2}. We plot the tangential velocity of a test particle in a Bose–Einstein condensate dark matter. The red circles with error bars show the observed data from the SPARC database \cite{Cd27}. The Newtonian gravitational contribution of the visible baryonic matter (stars and gas) is shown by the green dashed line. The blue diamonds show the velocity contribution of our fractional BEC dark matter core, excluding any baryonic contribution. Finally, the total rotation velocity $V_{\mathrm{tot}}$ given by (\ref{vtot}) is shown as the thick solid black line; we recall that this is the quadrature sum of the baryonic and dark matter components. 

Neither the isolated baryonic fraction (green) nor the BEC dark matter alone (blue) can reproduce the total tangential velocity (red), even with a significant baryonic component. Only the superposition of both distributions (black) yields a close match to the observational data. In systems such as UGC05918 and UGC08837, where baryons are important at the scales we are looking at, the BEC dark matter still provides the main gravitational potential at small scales (\(r < R_c\)), which causes the steep slope of the inner rotation curve. Only within a composite framework does the total velocity profile \(V_{\mathrm{tot}}\) correctly agree with observation, consistent with the fact that the BEC core continues to play a crucial dynamic role in baryon-rich environments.

\chg{Two remarks on the shape of the BEC contribution are in order. First, no CDM envelope enters $V_{\rm tot}$ (Eq.~\ref{vtot}), so the shape of the blue curves is determined by the condensate profile alone. Since the density has a finite, nearly constant central value (Fig.~\ref{dens}), the BEC velocity satisfies $V_{\rm tg}\simeq(4\pi G\rho_c/3)^{1/2}\,r$ at small radii, i.e.\ it rises linearly, as do the observed inner rotation curves. Second, this differs from the soliton--NFW models of Refs.~\cite{Bar18,Liu23,Cd31}, where the compact soliton produces a pronounced peak of the rotation curve at a radius comparable to the soliton radius, inside the range of the data. In our solutions the flat core extends over a region that is large compared with the radial range probed for most of the galaxies, the BEC velocity is still rising at the last measured point, and its maximum lies outside the data range. The absence of the peak therefore reflects the extended core of our solutions and not a different envelope model. 
We have not fitted the soliton--NFW profile to the present sample; a direct comparison with it, and with other cored profiles, is left for future work.}

\subsection{Sensitivity Analysis of the Model Parameters }
\label{sec:sens}

We note the concern that the degeneracy between the astrophysical parameters ($R_{c}$, $\Upsilon_{\text{disk}}$) could, in principle, permit acceptable fits for any chosen value of the particle mass $m$. We investigate this possibility by carrying out a global optimization over a subset of galaxies, holding the particle mass fixed at the Fuzzy Dark Matter (FDM) scale, $m = 10^{-22}$ eV$/c^2$, and a scattering length $a = 7.86 \times 10^{-79}\,{\rm cm}$, as predicted by the string axiverse model at the grand unification scale.

With the scaling relations in section 2.2, $\tilde{a} \propto 1/m^{2}$, increasing the particle mass (e.g., to $10^{-22}$ eV$/c^2$ or $10^{-15}$ eV$/c^2$) drives the interaction parameter $\tilde{a}$ toward zero, essentially reducing the model to non-interacting dark matter. Simultaneously, the characteristic mass scale $\alpha \equiv M(r)/\tilde{M}(\tilde{r}) \propto 1/m^{2}$ drops by several orders of magnitude. For a typical galactic core radius $R_{c} \sim 3$ kpc, a mass of $10^{-22}$ eV$/c^2$ gives a core mass of just $\sim 10^{2}\,M_{\odot}$, while a mass of $10^{-15}$ eV$/c^2$ leads to an almost negligible mass profile.
Importantly, the fitting procedure must specifically satisfy the constraint $R_{c} > R_{\text{disk}}$, which is strictly adhered to throughout the minimization process. One may wonder if a reasonable fit could still be obtained by freely adjusting $R_{c}$ for a larger mass. \chg{Since the mass-scaling factor $\alpha$ decreases as $m^{2}$, one would have to reduce $R_{c}$ by a comparable factor to compensate. This compensation is prevented by the condition $R_{c} > R_{\rm disk}$, which we impose in the fits; the test was not repeated without this constraint.} \chg{In our fits the product $MR_c$ is nearly the same for all galaxies (between 1.3 and $1.5\times10^{10}\,M_\odot$\,kpc in Table~\ref{table1}), as expected when the same dimensionless solution is used for every galaxy, so that $m$ enters only through the overall mass scale $\hbar^{2}/(2Gm^{2}R_c)$. Table~\ref{tab:sidebyside} shows the consequence of this scaling for $m=10^{-22}$ eV$/c^2$: with the same dimensionless solution and the constraint $R_c>R_{\rm disk}$, the core mass is too small by several orders of magnitude and the fits are poor.} The BEC dark matter model at the $10^{-26}$ eV$/c^2$ scale gives core masses of the right order and $\chi^{2} < 1$, whereas at $10^{-22}$ eV$/c^2$ the same procedure gives a strong suppression of the core mass and reduced chi-squared values between 6 and 23 for the nine galaxies of Table~\ref{tab:sidebyside}. \chg{This test illustrates the scaling of the solutions with $m$; it must not be read as an exclusion of $m=10^{-22}$ eV$/c^2$. Its outcome depends on the constraint $R_c>R_{\rm disk}$, which forces $R_c$ to values within $0.1$ kpc of $R_{\rm disk}$ in all nine galaxies (Table~\ref{tab:sidebyside}), and on the dimensionless solution used. For $m=10^{-22}$ eV$/c^2$ it is well known that self-gravitating solitons with kiloparsec-size cores and masses of $10^{8}$--$10^{9}\,M_\odot$ exist~\cite{Cd33}; they correspond to a much larger dimensionless mass than the one implied here.}

\begin{table}[h]
\centering
\caption{Side-by-side sensitivity comparison for a 9-galaxy subset. For both mass scales, the solver was granted freedom to optimize $R_{c}$ and $\Upsilon_{disk}$ subject to the physical constraint $R_c > R_{\text{disk}}$. }
\vspace{0.2cm}
{
}
\begin{tabular}{lc @{\hspace{0.5cm}} ccc @{\hspace{0.5cm}} ccc}
\toprule
 & \multicolumn{3}{c}{$m\approx 10^{-26}$ eV $/c^2$, $a\approx 10^{-82}$ cm} 
 & \multicolumn{3}{c}{$m\approx 10^{-22}$ eV $/c^2$, $a\approx 10^{-79}$ cm}\\
\cmidrule(lr){3-5} \cmidrule(l){6-8}\textbf{Galaxy} & \textbf{$R_{\text{disk}}$ (kpc)} & \textbf{$R_{c}$ (kpc)} & \textbf{$M$ ($10^{10}\,M_{\odot}$)} & \textbf{$\chi^2$} & \textbf{Constrained $R_{c}$}(kpc) & \textbf{($10^{10}\,M_{\odot}$)} & \textbf{$\chi^2$} \\
\midrule
KK98-251 & 1.340 & 5.297 & $0.27$ & 0.460 & 1.350 & $0.21 \times10^{-7}$ & 12.222 \\
NGC0100 & 1.660 & 5.195 & $0.28$ & 0.678 & 1.670 & $0.16\times10^{-7}$ & 12.390 \\
NGC3521 & 2.400 & 4.324 & $ 0.33$ & 0.180 & 2.500 & $0.11\times10^{-7}$ & 8.377\\
NGC4068 & 0.590 & 4.119  & $ 0.35$ & 0.178 & 0.600& $0.49\times10^{-7}$ & 6.326\\
NGC4183 & 2.790 &  7.027 & $0.21$ & 0.683 & 2.900 & $0.10\times10^{-7}$ & 15.610 \\
UGC04278  & 2.210 & 4.450 & $ 0.32$ &  0.397 & 2.220 & $ 0.13\times10^{-7}$ & 22.540\\
UGCA444 & 0.830 & 4.073 & $0.35$ & 0.922 & 0.840 & $ 0.35\times10^{-7}$ & 11.771 \\
UGC05005   & 3.200 & 7.494 & $0.20$ & 0.259 & 3.300 & $ 0.90\times10^{-8}$ & 10.072 \\
UGC08837   & 1.720 & 5.019 & $0.29$ & 0.325 & 1.730 & $ 0.17\times10^{-8}$ & 6.432 \\
\midrule
\bottomrule
\end{tabular}
\label{tab:sidebyside}
\end{table}

\begin{figure}[H]
\centering
\includegraphics[width=0.7\textwidth]{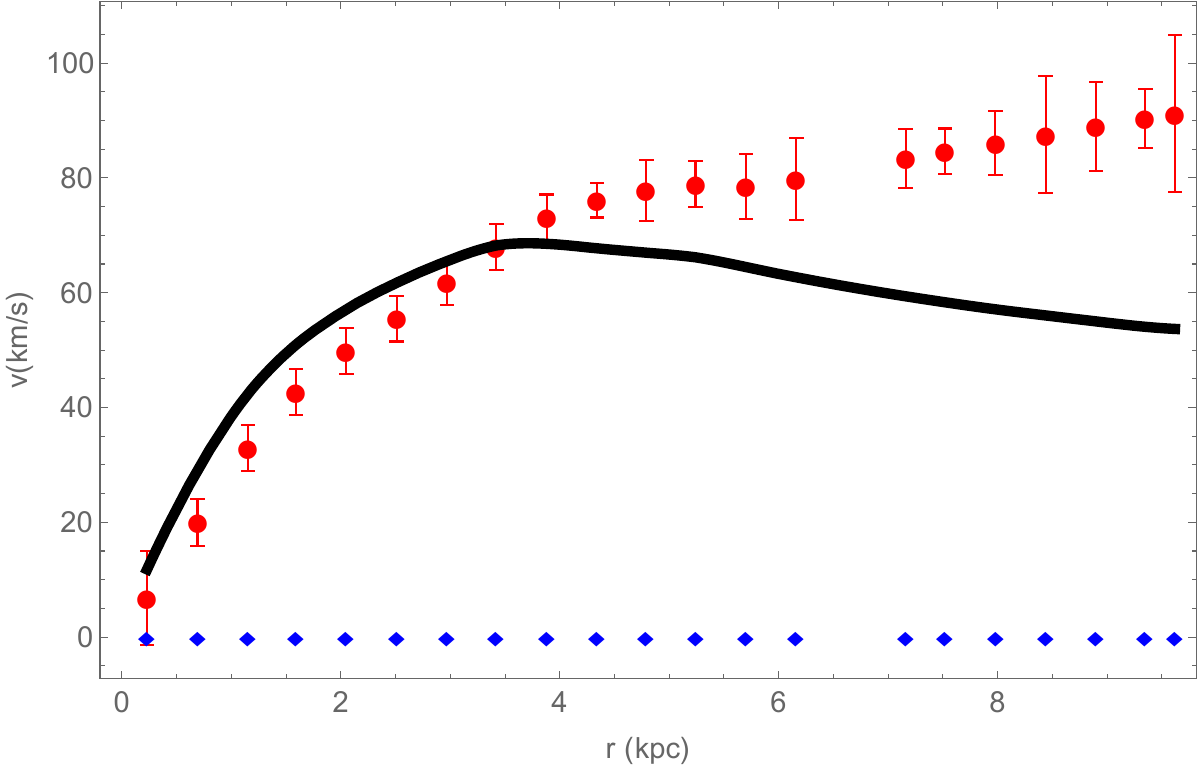}
\caption{Tangential velocity for NGC0100 at $m = 10^{-22}$ eV$/c^2$ and $a = 7.86\,10^{-79}$ cm. The modeled dark matter velocity (blue diamonds) flattens at $V_{\text{DM}} \approx 0$ as a result of the physical collapse of the core mass. The resulting total curve (black line), driven by the baryonic contribution alone, fails to reproduce the observed data (red).}
\label{figngc}
\end{figure}

As a direct illustration of this effect, Fig.~\ref{figngc} shows the
resulting rotation curve for NGC~0100 at $m = 10^{-22}$ eV$/c^2$. Since the core mass has effectively evaporated under this mass scale, the corresponding dark matter velocity contribution (blue diamonds) remains close to zero
across all radii.

\begin{figure}[H]
\centering
    \begin{subfigure}{0.45\textwidth}
        \includegraphics[width=\textwidth]{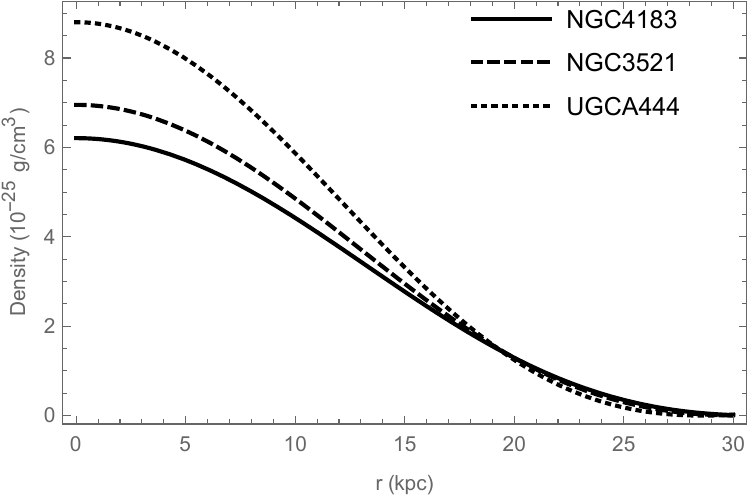}
\end{subfigure}
\caption{Density profiles \chg{of the condensate (BEC component only, GPP solution)} for NGC4183 (continuous), NGC3521 (dashed) and UGCA444 (dotted). \chg{The last SPARC data points of these galaxies lie at about 21, 6 and 2.6 kpc, respectively.}}
\label{dens}
\end{figure}

\chg{Finally, Fig.~\ref{dens} shows that the BEC component of the fractional model has a flat, finite-density core, in contrast with the central cusp predicted by standard CDM models \citep{Cd31,Cd33}. For three galaxies selected from our sample, the density of the condensate alone, obtained from the GPP solution (with no CDM envelope included), remains finite at the center and decreases smoothly outward, vanishing at the boundary of the computational domain. We verified that the same behavior holds for all galaxies in our sample. We do not claim, however, that this resolves the cusp--core problem: doing so would require comparison with other cored profiles and with CDM models that include baryonic feedback, which is beyond the scope of this work.}

\subsection{Necessary Stability Conditions}
\label{sec:stab}
For a self-gravitating quantum structure to exist, it is necessary that its total energy be negative \citep{Cd30}. We compute the total energy $E$ of the dark matter core for each galaxy in our sample to assess the physical viability of our solutions. Table \ref{table3} shows that for each galaxy studied, this total energy is negative, implying bound core structures within the fractional BEC framework.

It is worth clarifying the precise sense in which "stability" is used here. Negative total energy is a necessary condition for a bound equilibrium, and the fact that it holds across our entire sample is consistent with bound equilibrium configurations. 
This criterion, while encouraging, does not by itself guarantee dynamical stability against small perturbations: a rigorous assessment would require a linear stability analysis of the GPP system, tracking the growth or decay of small perturbations $\delta\psi(r,t)$ and $\delta V_{\rm grav}(r,t)$ around the equilibrium solution, or equivalently a time-dependent numerical evolution of the perturbed configuration. Such an analysis lies beyond the scope of the present work, but it represents a natural and promising next step, which we plan to pursue by extending our iterative relaxation framework into the time domain.

\begin{center}
\captionof{table}{Total energy of the dark matter core for each galaxy in the sample, computed using equation (\ref{eq8}).}
\begin{tabular}{lccccc}
\toprule
\textbf{Galaxy} & \multicolumn{1}{c}{$E(10^{54}J)$} & \textbf{Galaxy} & \multicolumn{1}{c}{$E(10^{54}J)$}\\
\midrule
D564-8   & $-7.81$   & UGC04278 & $-17.22$  \\
DDO064   & $-24.74$  & UGC04483 & $-21.99$   \\
KK98-251 & $-10.25$   & UGC05005 &$-3.65$    \\
NGC0100  & $-10.62$  & UGC05414 &$-14.22$   \\
NGC2976  & $-49.93$ & UGC05918 & $-2.18$\\
NGC3521  & $-18.68$ &   UGC07089 & $-7.10$ \\
NGC3949  & $-25.94$ &   UGC07232 & $-108.69$  \\
NGC3953  & $-3.25$  &  UGC07261 & $-10.07$    \\
NGC4068  & $-21.55$  &  UGC07323 & $-12.50$\\
NGC4088  & $-8.08$  & UGC07559 & $-12.09$  \\
NGC4183  & $-4.42$  & UGC07577 & $-6.28$    \\
NGC4389  & $-31.58$ & UGC07866 & $-10.32$    \\
UGC01281 & $-15.04$ & UGC08837 & $-12.02$  \\
UGC02023 & $-19.74$ & UGCA281  & $-26.73$    \\
UGC02455 & $-11.95$   & UGCA444  & $-22.28$   \\

\bottomrule
\end{tabular}
\label{table3}
\end{center}

\section{Comparison with Alternative Dark Matter Models}
\label{sec:discussion}
Our fractional BEC-DM model offers a \chg{phenomenological} alternative to the standard dark matter paradigms. Unlike the Navarro–Frenk–White (NFW) profile that predicts a steep central density cusp \citep{Cd34}, our model produces a cored density profile \chg{from the GPP system}. The numerical solutions we obtain give central densities in the range $\rho_{c} \approx 10^{-24}$--$10^{-26}$~g/cm$^3$. The finite-density core emerges naturally from the quantum nature of the dark matter particle, which regularizes the central cusp predicted by the standard $\Lambda$CDM models, a feature that has attracted growing attention in recent BEC-DM literature \citep{Cd31, Cd33}.
 
The NFW profile is the standard benchmark for large-scale cold dark matter distributions, but it is well known to overpredict central densities relative to observations of galactic cores. In contrast, our model generates flat cores from the ground up using the Gross–Pitaevskii–Poisson system, without the need for extra phenomenological assumptions. This also differs our approach from Self-Interacting Dark Matter (SIDM) models, where core formation usually relies on finely tuned scattering cross-sections \citep{Cd35}. Here, cores arise purely from basic quantum effects, enabling the fractional BEC-DM model to address the cusp-core tension \chg{with a small number of parameters ($m$ and $a$, and $R_c$ and $\Upsilon_{\rm disk}$ for each galaxy)}.
 
Although the SPARC rotation curves are reproduced with good accuracy, and the \chg{solutions have finite-density cores}, some limitations of the present approach should be mentioned. Specifically, even though numerical convergence is attained for $\tilde{r}_{max}=10$, the computational domain is still finite, in contrast to the physical galactic potential, which reaches out to infinitely large distances. We have thus verified the robustness of our results concerning the size of the numerical domain by repeating the calculations with cutoffs at $2\tilde{r}_{max}$ and $3\tilde{r}_{max}$. The profiles are essentially unchanged, even though extending the computational domain substantially increases the computational cost. This shows that, at the level of numerical accuracy considered here, our results are not affected by the specific choice of $\tilde{r}_{max}=10$. 

Second, the CDM background is considered as a fixed gravitational potential, not as a dynamically evolving component of the system. \chg{We recall that in the fits of Tables~\ref{table1}--\ref{table3} no CDM component is included in $V_{\rm tot}$ (Section~\ref{sec:density}); the test described here is a limited check.} We tested the sensitivity of our results to this assumption by repeating the fitting for three representative galaxies, NGC2976, UGC04483, and UGC07577, using two alternative descriptions of the outer halo: a standard NFW profile and an isothermal sphere. In both cases, the best-fit core radius $R_c$ changed by less than 15 \%, and the reduced $\chi^2$ remained below 1.0. These results indicate that the inferred properties of the inner region are fairly insensitive to the particular choice of outer halo profile. This can be seen from the fact that the latter influences mainly the dynamics at radii $r\gtrsim 2R_c$, while the inner core structure is set mostly by the balance involving the BEC quantum pressure. A more complete treatment would require the BEC and CDM components to be described as dynamically coupled rather than as a core embedded in a fixed background. Developing such fully coupled configurations, including possible multi-state solutions, is therefore an interesting direction for future work. \chg{Third, the baryons are not included in the condensate equation; this is justified by the estimate of Section~\ref{sec:baryons} for the very small boson mass used here but not for heavier bosons. Finally, the reduced chi-squared values below unity should be read with care: for most galaxies the number of data points is modest and they lie in the inner region, so that this statistic is not a stringent discriminator between cored profiles. A comparison with Burkert, pseudo-isothermal and soliton--NFW profiles using information criteria is left for future work.}


\section{Conclusion}
\label{sec:conclusion}


We solved the complete Gross–Pitaevskii–Poisson system for self-gravitating Bose–Einstein condensate dark matter halos, moving beyond the Thomas–Fermi approximation. This was achieved through a self-consistent iterative relaxation method that fully incorporates the kinetic (quantum-pressure) term within the mean-field framework. This gives convergent stationary configurations for ultralight bosons with $m \simeq 10^{-26}\,\mathrm{eV}/c^2$ and $a \simeq 10^{-82}\,\mathrm{cm}$, \chg{values whose order of magnitude is comparable to string-axiverse estimates at the GUT scale}. The condensate is taken to be a sub-dominant component of the halo, \chg{but we have not established its fraction: the ratio of the core mass to the halo mass is only illustrative for a single galaxy, and the cosmological and dynamical constraints on such a light component (Section~\ref{sec:constraints}) are not removed by the fractional assumption.}


The model, when faced with rotation curves from thirty SPARC dark-matter-dominated galaxies, attains $\chi^2 < 1$ by employing just two free parameters per galaxy ($\Upsilon_\mathrm{disk}$ and $R_c$), with the central density derived straight from the GPP equations instead of being independently fitted. The profiles thus obtained are flat and finite at the center. \chg{The data of most of these galaxies cover only the inner region with a modest number of points, so that $\chi^2<1$ indicates compatibility with the inner kinematics rather than a stringent test.}
All thirty condensate cores have a negative total energy \chg{as returned by the solver}, consistent with bound equilibrium; we interpret this as a necessary condition and not as a proof of dynamical (linear) stability, which is left for future work.

\chg{The sensitivity test shows that, with the same dimensionless solution and the constraint $R_c>R_{\rm disk}$, the choice $m=10^{-22}\,\mathrm{eV}/c^2$ gives $\chi^2\gg1$ in the nine galaxies tested; because this outcome depends on that constraint, we do not claim that the data determine $m$ and $a$. All fits of Table~\ref{table1} use a single choice of $(m,a)$ for the whole sample.}

The iterative GPP solver developed here, beyond its astrophysical application, is a general tool for self-gravitating and trapped Bose systems more broadly—boson stars, dilute condensates in compact objects, and large-$N$ trapped atomic condensates—where nonlinear mean-field interactions among a macroscopic but finite number of quanta govern the ground-state structure. Thus, the present halo problem is itself a case of a self-gravitating few/many-body system reduced to its mean-field limit, and the numerical strategy carries over directly to that broader class of problems. Future work will (i) extend the relaxation scheme to the time domain to assess linear dynamical stability, and (ii) perform a full Bayesian MCMC analysis over the complete SPARC sample, with systematic comparison to NFW, Burkert, soliton--NFW and Thomas–Fermi benchmarks\chg{, (iii) include the CDM envelope and the baryonic potential in the condensate equation, and (iv) assess the cosmological and dynamical constraints at this mass quantitatively}.

\section*{Funding Declaration}
This work is supported by the Ministry of Higher Education and Scientific Research, Algeria under the code: PRFU:B00L02UN020120220002. 

\section*{Competing Interest Declaration}
The authors declare that they have no competing interests.

\section*{Data Availability Statements}
\chg{The rotation-curve data analysed in this work are publicly available from the SPARC database~\cite{Cd27}. The fitted parameters are given in Tables~\ref{table1}--\ref{table3}. The numerical code is available from the corresponding author upon reasonable request.}

\end{document}